\documentclass[11pt,leqno]{article}
\usepackage{amsmath, amssymb}
\usepackage{latexsym}

\begin{document}

\newtheorem{thm}{Theorem}[section]
\newtheorem{cor}[thm]{Corollary}
\newtheorem{lem}[thm]{Lemma}
\newtheorem{prop}[thm]{Proposition}
\newtheorem{defn}[thm]{Definition}
\newtheorem{rem}[thm]{Remark}
\def\nm{\noalign{\medskip}}

\bibliographystyle{plain}

\numberwithin{equation}{section}

\newcommand{\qed}{\hfill \ensuremath{\square}}
\newcommand{\ds}{\displaystyle}
\newcommand{\pf}{\medskip \noindent {\sl Proof}. ~ }
\newcommand{\p}{\partial}
\renewcommand{\a}{\alpha}
\newcommand{\z}{\zeta}
\newcommand{\pd}[2]{\frac {\p #1}{\p #2}}
\newcommand{\norm}[1]{\| #1 \|}
\newcommand{\dbar}{\overline \p}
\newcommand{\eqnref}[1]{(\ref {#1})}
\newcommand{\na}{\nabla}
\newcommand{\Om}{\Omega}
\newcommand{\ep}{\epsilon}
\newcommand{\eps}{\varepsilon}
\newcommand{\vp}{\varphi}
\newcommand{\RR}{\mathbb{R}}
\newcommand{\CC}{\mathbb{C}}
\newcommand{\NN}{\mathbb{N}}
\newcommand{\nuu}{\tilde{\nu}}
\newcommand{\la}{\langle}
\newcommand{\ra}{\rangle}
\newcommand{\Scal}{\mathcal{S}}
\newcommand{\Kcal}{\mathcal{K}}
\newcommand{\Dcal}{\mathcal{D}}
\newcommand{\Pcal}{\mathcal{P}}
\newcommand{\Qcal}{\mathcal{Q}}

\title{High-Order Terms in the Asymptotic Expansions of the Steady-State
Voltage Potentials in the Presence of Conductivity Inhomogeneities
of Small Diameter \thanks{This work was completed while both
authors were at the Mathematical Sciences Research Institute
(MSRI) during the special program on inverse problems. They would
like to thank Gunther Uhlmann for invitation and the Institute for
partial support and providing stimulating environment.} }

\author{Habib Ammari \thanks{partly supported by ACI Jeunes Chercheurs (0693)
from the Ministry of Education and Scientific Research, France.}\\
 Centre de Math\'ematiques Appliqu\'ees\\
Ecole Polytechnique \\ 91128 Palaiseau Cedex, France
\\ ammari@cmapx.polytechnique.fr \and Hyeonbae Kang\thanks{partly
supported by KOSEF 98-0701-03-5 and BK21 at the School of Mathematical Sciences of SNU.} \\
School of Mathematical Sciences \\ Seoul National University \\
Seoul 151-747, Korea \\ hkang@math.snu.ac.kr}

\maketitle

\begin{abstract}
We derive high-order terms in the asymptotic expansions of the
steady-state voltage potentials in the presence of a finite number
of diametrically small inhomogeneities with conductivities
different from the background conductivity. Our derivation is
rigorous, and based on layer potential techniques. The asymptotic
expansions in this paper are valid for inhomogeneities with
Lipschitz boundaries and those with extreme conductivities.
\end{abstract}

\noindent {\footnotesize Mathematics subject classification
(MSC2000): 35B30}


\noindent {\footnotesize Short title: Small conductivity
inhomogeneities}


\section{Introduction}


Let $\Om$ be a bounded domain in $\RR^d$, $d \ge 2$, with a
connected Lipschitz boundary $\partial \Omega$. Let $\nu$ denote
the unit outward normal to $\partial \Omega$. Suppose that $\Om$
contains a finite number $m$ of small inhomogeneities
$(D_l)_{l=1}^m$, each  of the form $D_l =z_l + \epsilon B_l$,
where $B_l$, $l=1, \ldots, m$, is a bounded Lipschitz domain in
$\RR^d$ containing the origin. We assume that the domains
$(D_l)_{l=1}^m$ are separated apart from each other and apart from
the boundary. More precisely, we assume that there exists a
constant $c_0 >0$ such that
\begin{equation}
\label{f1}
 | z_l - z_{l^\prime}  | \geq 2 c_0 > 0, \forall \; l \neq l^\prime  \quad
\mbox{ and } \mbox{ dist} (z_l, \partial \Omega) \geq 2 c_0
> 0, \forall \; l,
\end{equation} and $\epsilon$, the common order of magnitude of the diameters
of the inhomogeneities, is sufficiently small, that these
 inhomogeneities are disjoint,  and that
their distance to $\RR^d \setminus \overline{\Omega}$ is larger
than $c_0$. We also assume that the "background" is homogeneous
with conductivity $1$ and the inhomogeneities $D_l$ have
conductivities $k_l$, $k_l \neq 1$, $1 \leq l \leq m$.

Let $u_\ep$ denote the steady-state voltage potential in the
presence of the conductivity inhomogeneities, {\em i.e.}, the
solution to
\begin{equation} \label{nepsg}
 \ \left \{
\begin{array}{l}
\ds \nabla  \cdot \Bigr( \chi(\Omega \setminus \bigcup_{l=1}^m
\overline{D_l}) + \sum_{l=1}^m k_l \chi(D_l) \Bigr) \nabla
u_\epsilon =0 \quad \mbox{in } \Om, \\ \nm \ds
\pd{u_\epsilon}{\nu} \big|_{\p\Om} =g.
\end{array}
\right .
\end{equation}

Let $U$ denote the "background" potential, that is, the solution
to \begin{equation} \label{ng}
\begin{cases}
\Delta U=0 \quad \mbox{in } \Om, \\ \nm \ds \pd{U}{\nu}
\big|_{\p\Om} =g.
\end{cases}
\end{equation}
 The function $g$ represents the applied boundary current; it
belongs to $ L^2_0(\p\Om) =\{ g \in  L^2(\p\Om), \int_{\p \Om} g =
0\}$. The potentials, $u_\ep$ and $U$, are normalized by $\int_{\p
\Om} u_\ep = \int_{\p \Om} U =0$.

The main achievement of this paper is a rigorous derivation, based
on layer potential techniques, of high-order terms in the
asymptotic expansion of $u_\ep |_{\p \Om}$ as $\ep \rightarrow 0$.
The leading order term in this asymptotic formula has been derived
by Cedio-Fengya et al. \cite{CMV98}; see also the prior work of
Friedman and Vogelius \cite{FV89} for the case of perfectly
conducting or insulating inhomogeneities. The main result of this
paper is the following full asymptotic expansion of the solution
for the case $m=1$:

\begin{thm} \label{thm:main}
Suppose that the inhomogeneity consist of single component and let
$u_\eps$ be the solution of \eqnref{nepsg}. The following pointwise
asymptotic expansion on $\partial \Omega$ holds for $d=2,3$:
\begin{equation} \label{final3}
\begin{array}{ll}
\ds u_\eps(x) &= U(x)  - \ds  \ep^{d-2} \sum_{|i| =1}^n
\sum_{|j|=1}^{n-|i| +1} \frac{1}{j!} \epsilon^{|i| + |j|} \times \\
\nm & \quad \quad \ds  \Bigr[
\ds \Bigr( (I + \sum_{p=1}^{n +2 - |i| -|j| -d} \epsilon^{d + p-1}
  \Qcal_p) (\partial^l U(z)) \Bigr)_i  M_{ij} \p^j_z N(x,z) \Bigr]
\\ \nm & \quad +  \ds  O(\ep^{d+n}),
\end{array}
\end{equation}
where the remainder $ O(\ep^{d+n})$ is dominated by $C \ep^{d+n} \|
 g \|_{L^2 (\p\Om)}$ for some $C$ independent of $x\in
\p\Om$. Here $N(x, z)$ is the Neumann function, $M_{ij}$, $i,j \in
\NN^d$,  are the generalized polarization tensors defined in
(\ref{mij}), and the operator $\Qcal_p$ is defined in
\eqnref{qcal}.
\end{thm}

We have a similar expansion for the solutions of the Dirichlet problem
(Theorem \ref{thm:main2}.

The derivation of the asymptotic expansions for any fixed number
$m$ of well separated inhomogeneities (these are a fixed distance
apart) follows by iteration of the arguments that we will present
for the case $m=1$. In other words, we may develop asymptotic
formulas involving the difference between the fields $u_\epsilon$
and $U$ on $\partial \Omega$ with $l$ inhomogeneities and those
with $l-1$ inhomogeneities, $l=m, \ldots, 1,$ and then at the end
essentially form the sum of these $m$ formulas (the reference
fields change, but that may easily remedied). The derivation of
each of the $m$ formulas is virtually identical.

We also note that the asymptotic expansion \eqnref{final3} is
valid for inhomogeneities with zero or infinity conductivity
(cavity or perfect conductor). Precise definitions of generalized
polarization tensors (GPT) associated with the domains $B_l$ and
the conductivities $k_l$ will be given at the end of section 3.
These GPT seem to be natural generalizations of the tensors that
have been introduced by Schiffer and Szeg\"{o} \cite{SS49} and
thoroughly studied by many other authors \cite{PS51}, \cite{KS86},
\cite{FV89}, \cite{CMV98}. (See section 3.)

The higher-order terms are essential when $\nabla U(z_l) =0$, for
then the leading order term in the asymptotic expansion of $u_\ep
|_{\p \Om}$, given in \cite{CMV98}, vanishes. We remind the reader
that, for general current inputs $g$,  $\nabla U$ vanishes at some
"critical points" inside $\Omega$.

The proof of our asymptotic expansion is radically different from
the ones in \cite{FV89}, \cite{CMV98}, and \cite{VV00}. It is
based on layer potential techniques and a decomposition formula of
the steady-state voltage potential into a harmonic part and a
refraction part. This formula is due to Kang and Seo \cite{KS96}.
What makes our proof particularly original and elegant is that the
rigorous derivation of high-order terms follows almost
immediately. The extension of the techniques used in  \cite{FV89},
\cite{CMV98}, and \cite{VV00} to construct higher-order terms in
the expansion of $u_\ep |_{\p \Om}$ as $\ep \rightarrow 0$ seems
to be laborious. Furthermore, the general approach developed in
this paper could be carried out to obtain more precise asymptotic
formulas for the full Maxwell's equations and for the equations of
linear elasticity than those derived in \cite{AVV01} and
\cite{AA02}. The method of this paper also enables us to extend
the asymptotic expansions to the cases of inhomogeneities with
Lipschitz boundaries. Previously, the leading order term was
derived under the assumption that inhomogeneities are $C^{1,
\alpha}$ smooth \cite{FV89} \cite{CMV98}. We note that our method
works as well even when the inhomogeneities have extreme
conductivities ($k=0$ or $k = \infty$).

Let us now explain what makes this asymptotic formula interesting
in the electrical impedance tomography (EIT). It is well-known
that the ultimate objective of EIT is to  recover, most
efficiently and accurately, the conductivity distribution inside a
body from measurements of current flows and voltages on the body's
surface. The vast and growing literature reflects the many
possible applications of EIT, e.g. for medical diagnosis or
nondestructive evaluation of materials \cite{CIN99}. In its most
general form EIT is severely ill-posed and nonlinear. Taking
advantage of the smallness of the inhomogeneities, Cedio-Fengya et
al. \cite{CMV98} have used  the leading order term in the
asymptotic expansion of $u_\ep |_{\p \Om}$ to find the locations
$z_l$, $l=1, \ldots, m$ of the inhomogeneities and certain
properties of the domains $B_l$, $l=1, \ldots, m$ (relative size,
orientation). The algorithm proposed by Cedio-Fengya et al.
\cite{CMV98} is based on a least-square algorithm. Ammari et al.
\cite{AMV00} have also utilized this leading order term to design
a variationally based direct reconstruction method. The new idea
in \cite{AMV00} is to form the integral of the "measured boundary
data" against harmonic test functions and choose the input current
$g$ so as to obtain expression involving the inverse Fourier
transform of distributions supported at the locations $z_l$, $l=1,
\ldots, m$. Applying a direct Fourier transform to this data then
pins down the locations. This approach is similar to the ideas
used by Calder\'{o}n \cite{C} in his proof of uniqueness of the
linearized conductivity problem and later, by Sylvester and
Uhlmann in their important work \cite{SU} on uniqueness of the
three-dimensional inverse conductivity problem. An other algorithm
that makes use of an asymptotic expansion of the voltage
potentials has been derived by Br\"{u}hl et al. \cite{BHV01}. This
algorithm is in the spirit of the linear sampling method of Colton
and Kirsch \cite{CK96}.

In all of these algorithms, the
 locations $z_l$, $l=1, \ldots, m$ of the inhomogeneities are found
 with an error $O(\ep)$ and little about the domains $B_l$ can be
 reconstructed. Making use of higher-order terms in the asymptotic
expansion of $u_\ep |_{\p \Om}$  we certainly would be able to
reconstruct the small inhomogeneities with higher resolution from
boundary information about specific solutions to (\ref{nepsg}).
Perhaps, more importantly, this would allow us to identify quite
general conductivity inhomogeneities without restrictions on their
sizes.

The use of higher-order terms in the asymptotic expansion of
$u_\ep |_{\p \Om}$ may also be decisive in dramatically improving
the algorithm of Kwon et al. \cite{KSY01}, that is based on the
observation of the pattern of a simple weighted combination of an
input current $g$ of the form $g = a \cdot \nu$ for some constant
vector $a$ and the corresponding output voltage. We also believe
that the use of such higher-order terms would improve the
algorithm of Mast et al. \cite{MNW97}, that uses eigenfunctions of
the scattering operator.

This paper is organized as follows. In section 2, we collect some
notations and preliminary regarding layer potentials. In section
3, we introduce the generalized polarization tensors associated
with the domains $D_l$ and the conductivities $k_l$. In section 4,
we provide a rigorous derivation of high-order terms in the
asymptotic expansion of the output voltage potentials.  For
reasons of brevity we restrict a significant part of this
derivation to the case of a single inhomogeneity $(m=1)$. The
proof in the case of multiple well-separated inhomogeneities may
be derived by a fairly straightforward iteration of the arguments
we present, however we leave the details to the reader.


\section{Layer Potentials for the Laplacian}


Let us first review some well-known properties of the layer
potentials for the Laplacian and prove some useful identities.

The theory of layer potentials has been developed in relation to
the boundary value problems. Let $D$ be a bounded domain in
$\RR^d, d \ge 2$. We assume that $\p D$ is Lipschitz. Let $\Gamma
(x)$ be the fundamental solution of the Laplacian $\Delta$:
\begin{equation}
\Gamma (x) =
\begin{cases}
\ds \frac{1}{2\pi} \ln |x|, \qquad & d=2, \\ \nm \ds
\frac{1}{(2-d)\omega_d} |x|^{2-d}, \qquad & d \ge 3,
\end{cases}
\end{equation}
where $\omega_d$ is the area of $(d-1)$ dimensional unit sphere.
The single and double layer potentials of the density function
$\phi$ on $D$ is defined by
\begin{align}
\Scal_D \phi (x) &:= \int_{\p D} \Gamma (x-y) \phi (y) d \sigma
(y) , \quad x \in \RR^d, \label{single}\\
\nm  \Dcal_D \phi (x) &:= \int_{\p D}
\pd{}{\nu_y}\Gamma (x-y) \phi (y) d \sigma (y) , \quad x \in \RR^d
\setminus \p D. \label{double}
\end{align}
For a function $u$ defined on $\RR^d \setminus \p D$, we denote
$$
\pd{}{\nu^\pm} u(x) := \lim_{t \to 0^+} \la \nabla u(x \pm t
\nu_x), \nu_x \ra, \quad x \in \p D
$$
if the limit exists. Here $\nu_x$ is the outward unit normal to
$\p \Om$ at $x$.

The proof of the following trace formula can be found in
\cite{FJR78, Folland76, N88} (for Lipschitz domains, see
\cite{Verchota84}):
\begin{align}
\pd{}{\nu^\pm} \Scal_D \phi(x) &= (\pm \frac{1}{2} I + \Kcal_D^*)
\phi (x),
\label{singlejump}\\
(\Dcal \phi)|_{\pm} & = (\mp \frac{1}{2} I + \Kcal_D) \phi (x),
\quad x \in \p D, \label{doublejump}
\end{align}
where $$ \Kcal _D \phi (x) = \frac{1}{\omega_d} \mbox{p.v.}
\int_{\p D} \frac{\la x-y, \nu_y \ra}{|x-y|^d} \phi(y) d \sigma(y)
$$ and $\Kcal_D^*$ is the $L^2$-adjoint of $\Kcal_D$. When $\p D$
is Lipschitz, $\Kcal_D$ is a singular integral operator and known
to be bounded on $L^2 (\p\Om)$ \cite{CMM82}. Let $L^2_0 (\p D) : =
\{ f \in L^2(\p D) : \int_{\p D} f d\sigma=0 \}$. The following
results are due to Verchota and Escauriaza-Fabes-Verchota.

\begin{thm}[\cite{EFV92}, \cite{Verchota84}] \label{Verchota}
$\lambda I - \Kcal_D^*$ is invertible on $L^\infty_0 (\p D)$ if
$|\lambda| \ge \frac{1}{2}$, and for $\lambda \in (-\infty,
-\frac{1}{2}] \cup (\frac{1}{2}, \infty )$, $\lambda I -
\Kcal_D^*$ is invertible on $L^2 (\p D)$.
\end{thm}

For proofs when $\p D$ is smooth, see \cite{FJR78},
\cite{Folland76}.

\medskip

The following theorem was proved in \cite{KS96, KS99, KS2000}.
\begin{thm}
Suppose that $D$ is a domain compactly contained in $\Omega$ with
a connected Lipschitz boundary and conductivity $k$. Then the
solution $u$ of the problem
\begin{equation} \label{ud}
 \ \left \{
\begin{array}{l}
\ds \nabla  \cdot ( (1 + (k-1) \chi(D)) \nabla u)=0 \quad \mbox{in
} \Om,
\\ \nm \ds \pd{u}{\nu} \big|_{\p\Om} =g
\end{array}
\right .
\end{equation}
is represented as
\begin{equation} \label{KS}
u(x) = H(x) + \Scal_D \phi (x), \quad x \in \Om,
\end{equation}
where the harmonic function $H$ is given by
\begin{equation} \label{KS-1}
H(x) = -\Scal_\Omega( g)(x) + \Dcal_\Omega (f)(x), \quad x \in
\Om, \quad f := u|_{\p \Omega},
\end{equation}
and $\phi \in L^2_0(\partial D)$ satisfies the integral equation
\begin{equation} \label{KS-2}
(\frac{k+1}{2(k-1)} I - \Kcal_D^* ) \phi = \pd{H}{\nu} \big |_{\p
D} \quad \mbox{on } \p D.
\end{equation}
Moreover, $\forall \; n \in \NN$, there exists a constant
$C_n=C(n, \Omega, \mbox{dist}(D, \partial \Omega))$ independent of
$|D|$ and $k$ such that
\begin{equation} \label{KS-3}
 \norm{H}_{\mathcal{C}^n(\overline{D})} \le C_n
\norm{g}_{L^2(\p \Om)}.
\end{equation}
\end{thm}

\pf The representation formula \eqnref{KS} was proved in
\cite{KS96, KS2000}. \eqnref{KS-3} was proved in \cite{KS99} for
$d=2$ and it is easily seen that the same proof works for $d=3$.
We only need to check carefully whether the constant $C_n$ in the
estimate \eqnref{KS-3} is independent of $|D|$. Before doing this,
let us point out that the harmonic function $H$ can be computed
explicitly from the boundary measurements $(\pd{u}{\nu}
\big|_{\p\Om}, u\big|_{\p\Om})$ and the density $\phi$ is uniquely
and explicitly determined by the domain $D$ and the harmonic
function $H$. The decomposition of the function $u$ into a
harmonic part $H$ and a refraction part $\Scal_D \phi$ is unique
\cite{KS96, KS2000}.  The representation formula \eqnref{KS} seems
to inherit geometric properties of $D$.

Suppose that $\mbox{dist}(D,
\partial \Omega)) > 2 c_0$ for some constant
$c_0 > 0$. From the definition  of $H$ in \eqnref{KS-1} it is easy
to
 see that
\begin{equation}
\label{h1} \norm{H}_{\mathcal{C}^n(\overline{D})} \le C_n \Bigr(
\norm{g}_{L^2(\p \Om)} + \norm{u|_{\p \Omega} }_{L^2(\p \Om)}
\Bigr),
\end{equation}
where $C_n$ depends only on $n$, $\p \Omega$, and $c_0$. Let
$\vec{\alpha}$ be a vector field supported in the set $\mbox{dist}
(x, \p \Om) < 2 c_0$ such that $\vec{\alpha} \cdot \nu(x) \ge
\delta$ for some $\delta >0$ for all $x \in \p\Om$. Using the
Rellich identity with this $\vec{\alpha}$, we can show that $$ \|
\pd{u}{T} \|_{L^2 (\p\Om)} \le C \Bigr( \| g \|_{L^2 (\p\Om)} + \|
\nabla u \|_{L^2 (\Om \setminus \overline D)} \Bigr), $$ where $C$
depends only on $\p\Om$ and $c_0$ and $T(x)$ is the tangent vector
to $\partial \Omega$ at $x$. (See the proof of Lemma 2.1 of
\cite{FKS99} for the detail of the proof.) Observe that
\begin{align*}
\norm{\nabla u }_{L^2(\Om \setminus \overline D)}^2 & \leq C
\int_\Omega (1 + (k-1)
\chi(D)) \nabla u \cdot \nabla u \; dx \\
& = C \int_{\partial \Omega} g u \; d\sigma \\
& \leq C \norm{g}_{L^2(\p \Om)} \norm{u|_{\p \Omega}}_{L^2(\p
\Om)}.
\end{align*}
Since $\int_{\partial \Omega} u \; d\sigma = 0,$ it follows from
the Poincar\'e inequality that
$$
\norm{u|_{\p \Omega}}_{L^2(\p \Om)} \leq C \| \pd{u}{T} \|_{L^2
(\p\Om)}.
$$
Thus we obtain
\begin{equation*}
\norm{u|_{\p \Omega} }_{L^2(\p \Om)}^2 \leq C \Bigr(
\norm{g}_{L^2(\p \Om)}^2 + \norm{g}_{L^2(\p \Om)} \norm{u|_{\p
\Omega}}_{L^2(\p \Om)} \Bigr) ,
\end{equation*}
and hence
\begin{equation*}
\norm{u|_{\p \Omega} }_{L^2(\p \Om)} \leq C  \norm{g}_{L^2(\p
\Om)} .
\end{equation*}
From (\ref{h1}) we finally obtain \eqnref{KS-3}. \qed

Using above representation we can derive a formula similar to
(\ref{final3}) which is potentially useful in detecting the
inhomogeneities (see the remark at the end of this paper). However
it uses the function $H$ which depends on $D$ and hence on $\ep$.
Thus in order to derive (\ref{final3}) we will transform it to
representations using only background potentials.

Let $N(x,y)$ be the Neumann function for $\Delta$ in $\Om$
corresponding to a Dirac mass at $z$. That is, $N$ is the solution
to
$$ \ \left \{
\begin{array}{l}
\ds \Delta_x N(x, z) = - \delta_z \quad \mbox{in } \Om, \\ \nm \ds
\pd{N}{\nu} \big|_{\p\Om} = - \frac{1}{|\partial \Omega|}.
\end{array}
\right .
$$
In addition, we assume that
\begin{equation} \label{Neumann-1}
\int_{\p\Om} N(x,y) d \sigma(x) =0 \quad \mbox{for } y \in \Om.
\end{equation}
Let us fix one more notation: For $D$, a subset of $\Om$, let
$$
N_D f (x) : = \int_{\p D} N(x,y) f (y) d \sigma (y).
$$

The following lemma relates the fundamental solution with the
Neumann function.

\begin{lem} \label{Neuman}
For $z \in \Om$ and $x \in \p\Om$, let $\Gamma_z(x) : =
\Gamma(x-z)$ and $N_z(x) := N(x,z)$. Then
\begin{equation} \label{Neumann-2}
(-\frac{1}{2} I + \Kcal_\Om) (N_z)(x) = \Gamma_z (x) \quad
\mbox{modulo constants}, \quad x \in \p\Om,
\end{equation}
or to be more precise, for any simply connected Lipschitz domain
$D$ compactly contained in $\Om$ and for any $g \in L^2_0 (\p D)$,
we have
\begin{equation} \label{Neumann-3}
\int_{\p D} (-\frac{1}{2} I + \Kcal_\Om) (N_z)(x) g(z) d\sigma (z)
= \int_{\p D} \Gamma_z (x) g (z) d\sigma(z), \quad \forall x \in
\p\Om.
\end{equation}
\end{lem}

\pf Let $f \in L^2_0(\p \Om)$ and define $$ u(z) := \la
(-\frac{1}{2} I + \Kcal_\Om) (N_z) , f \ra_{\p \Om} \quad z \in
\Om. $$ Then $$ u(z) = \int_{\p\Om} N(x,z) (-\frac{1}{2} I +
\Kcal_\Om^* ) f(z) d \sigma(z). $$ Therefore, $\Delta u=0$ in
$\Om$ and $\ds \pd{u}{\nu}|_{\p\Om} = (-\frac{1}{2} I +
\Kcal_\Om^* ) f$. Thus by \eqnref{singlejump} we have $$ u (z) -
\Scal_\Om f (z) = \mbox{constant}, \quad z \in \Om. $$ Thus if $g
\in L^2_0(\p\Om)$, then we obtain \[\begin{array}{l} \ds
\int_{\p\Om} \int_{\p D} (-\frac{1}{2} I + \Kcal_\Om) (N_z)(x)
g(z) d\sigma(z) f(x) d \sigma(x) \\ \nm \quad \ds = \int_{\p\Om}
\int_{\p D} \Gamma_z (x) g(z) d\sigma(z) f(x) d \sigma(x).
\end{array}\] Since $f$ is arbitrary, we have \eqnref{Neumann-2} or
equivalently, \eqnref{Neumann-3}. This completes the proof.
 \qed

\medskip
Let $g \in L^2_0(\p \Om)$. Let $U(y) := \int_{\p\Om} N(x,y) g(x) d
\sigma(x)$. Then $U$ satisfies
\begin{equation} \label{back}
\begin{cases}
& \ds \Delta U =0 \quad \mbox{in } \Om,  \\
\nm &  \ds \pd{U}{\nu}|_{\p\Om} = g \in L^2_0 (\p \Om) , \\
\nm & \ds  \int_{\p\Om} U(x) d\sigma(x) = 0 .
\end{cases}
\end{equation}

\begin{thm} \label{rep-n}
The solution $u$ of \eqnref{ud} can be represented as
\begin{equation} \label{rep-n-1}
u(x) = U(x) - N_D \phi (x), \quad x \in \p\Om
\end{equation}
where $\phi$ is defined in \eqnref{KS-2}.
\end{thm}

\pf By substituting \eqnref{KS} into the equation \eqnref{KS-1},
we obtain
$$
H(x) = -\Scal_\Omega( g)(x) + \Dcal_\Omega (H|_{\p\Om} + (\Scal_D
\phi)|_{\p\Om} )(x) , \quad x \in \Om.
$$
It then follows from \eqnref{doublejump} that
\begin{equation} \label{rep-n-2}
(\frac{1}{2} I - \Kcal_\Om ) (H|_{\p \Om}) = -( \Scal_\Om g)|_{\p
\Om} + (\frac{1}{2} I + \Kcal_\Om ) ( (\Scal_D \phi)|_{\p\Om} ) ,
\quad \mbox{on } \p \Om.
\end{equation}
Since $U = -\Scal_\Om( g) + \Dcal_\Om (U|_{\p\Om})$ in $\Om$, we
have
\begin{equation} \label{rep-n-3}
(\frac{1}{2} I - \Kcal_\Om ) (U|_{\p \Om}) = -( \Scal_\Om g)|_{\p
\Om}.
\end{equation}
Since $\phi \in L^2_0 (\p D)$, it follows from \eqnref{Neumann-2}
that
\begin{equation} \label{rep-n-4}
- (\frac{1}{2} I - \Kcal_\Om ) ((N_D \phi)|_{\p \Om})= (\Scal_D
\phi)|_{\p\Om}.
\end{equation}
From \eqnref{rep-n-2}, \eqnref{rep-n-3}, and \eqnref{rep-n-4}, we
conclude that
\begin{equation*}
(\frac{1}{2} I - \Kcal_\Om ) ( H|_{\p \Om} - U|_{\p \Om} +
(\frac{1}{2} I + \Kcal_\Om ) ((N_D \phi)|_{\p \Om}) ) = 0.
\end{equation*}
Therefore, we have
\begin{equation} \label{rep-n-5}
H|_{\p \Om} - U|_{\p \Om} + (\frac{1}{2} I + \Kcal_\Om ) ((N_D
\phi)|_{\p \Om}) = C \ (\mbox{constant}).
\end{equation}
Note that $(\frac{1}{2} I + \Kcal_\Om ) ((N_D \phi)|_{\p \Om})=
(N_D \phi)|_{\p \Om} + (\Scal_D \phi)|_{\p\Om}$. Thus we get from
\eqnref{KS} and \eqnref{rep-n-5}
\begin{equation} \label{rep-n-6}
u|_{\p\Om} = U|_{\p \Om} - (N_D \phi)|_{\p \Om} + C.
\end{equation}
Since all the functions entering in \eqnref{rep-n-6} belong to
$L^2_0(\p\Om)$, we conclude that $C=0$, and the theorem is proved.
\qed

\medskip

We have a similar representation for solutions of the Dirichlet
problem. Let $G(x,y)$ be the Green function for the Dirichlet
problem, i.e., the function $V$ defined by $V(x) := \int_{\p\Om}
\frac{\partial G}{\partial \nu(y)} (x,y) f(y) d\sigma (y)$ is the
solution of the problem $\Delta V=0$ in $\Om$ and $V|_{\p\Om}=f$
for any $f \in L^2 (\p\Om)$. Then we have the following
representation theorem.

\begin{thm} \label{rep-d}
\begin{equation} \label{Diri-2}
(\frac{1}{2} I + \Kcal_\Om^*)^{-1} (\frac{\partial
\Gamma_z(y)}{\partial \nu(y)} )(x) = \frac{\partial G_z}{\partial
\nu(x)}  (x) , \quad x \in \p\Om, \ z \in \Om.
\end{equation}
Let $u$ be the solution of \eqnref{ud} with the Neumann condition
replaced by the Dirichlet condition $u|_{\p\Om}=f$. Then $u$ can
be represented as
\begin{equation} \label{rep-d-1}
\pd{u}{\nu} (x) = \pd{V}{\nu} (x) - G_D \phi (x), \quad x \in
\p\Om
\end{equation}
where $\phi$ is defined in \eqnref{KS-2} and $\ds G_D \phi (x):=
\int_{\p D}\frac{\partial  G}{\partial \nu(y)} (x,y) \phi(y) d
\sigma (y)$.
\end{thm}

Theorem \ref{rep-d} can be proved in the same way as Theorem
\ref{rep-n}. In fact, it is simpler because of the solvability of
the Dirichlet problem, or equivalently, the invertibility of
$(\frac{1}{2} I + \Kcal_\Om^*)$. So we omit the proof.


\section{Generalized Polarization Tensors}


In this section we introduce the generalized polarization tensors
(GPT) associated with a domain $B$ and a conductivity $k$. These
GPT are the basic building block for the asymptotic expansions in
this paper.

Let $B$ be a Lipschitz bounded domain in $\RR^d$ and the
conductivity of $B$ be $k$ $(k \neq 1)$. The polarization tensor
is $M=(m_{ij})$, $1 \leq i, j \leq d$, is defined by
$$
m_{ij}:= (k -1) \left [ \delta_{ij} |B| + \int_{\p B} y_i
\pd{}{\nu^+} \psi_j (y) d \sigma (y) \right ],
$$
where $\psi_j$ is the unique solution of the following
transmission problem:
$$
\begin{cases}
\Delta \psi_j (x) = 0, \quad x \in B \cup \RR^d \setminus
\overline B, \\
\nm \ds \psi_j |_+ - \psi_j |_- =0 \quad \mbox{on } \p B,
\\
\nm\ds \pd{}{\nu^+} \psi_j  - k \pd{}{\nu^-} \psi_j  = \nu_j \quad
\mbox{on } \p B, \\
\nm \ds \psi_j (x) \to 0 \mbox{ as } |x| \to \infty.
\end{cases}
$$
See \cite{SS49}, \cite{CMV98}, and \cite{FV89}. One can easily
check using \eqnref{singlejump} that
$$
\psi_j = \Scal_{B} (\lambda I - \Kcal_{B}^*)^{-1}(\nu_j).
$$
Using \eqnref{singlejump} again, we have
\begin{align*}
\int_{\p B} & y_i \pd{}{\nu^+} \psi_j (y) d \sigma (y) \\
 & = -
\int_{\p B} y_i (\frac{1}{2} I - \Kcal_{B}^*) (\lambda I -
\Kcal_{B}^*)^{-1}(\nu_j) (y) d \sigma (y) \\ & = - \int_{\p B} y_i
\nu_j d \sigma (y) + (\lambda - \frac{1}{2}) \int_{\p B} y_i
(\lambda I - \Kcal_{B}^*)^{-1} (\nu_j) (y) d \sigma (y) \\ & = -
\delta_{ij} |B| + \frac{1}{k-1}  \int_{\p B} y_i (\lambda I -
\Kcal_{B}^*)^{-1} (\nu_j) (y) d \sigma (y).
\end{align*}
Therefore we prove that the polarization tensor $M$ associated
with $B$ and $k$ is given by
\begin{equation} \label{polar}
m_{ij} = \int_{\p B} y_i (\lambda I - \Kcal_{B}^*)^{-1}(\nu_j) (y)
d \sigma (y) .
\end{equation}
Recall $\lambda := \frac{k+1}{2(k-1)}$.

For a multi-index $i=(i_1, \ldots, i_d) \in \NN^d$, let $\p^i f =
\p^{i_1}_1 \ldots \p^{i_d}_d f$ and $x^i := x_1^{i_1} \cdots
x_d^{i_d}$. For $i, j \in \NN^d$, we define the {\it generalized
polarization tensor} $M_{ij}$ by
\begin{equation}
\label{mij} \ds M_{ij} := \int_{\p B} y^j  \phi_i (y) d \sigma(y),
\end{equation}
where $\phi_i$ is defined by $$ \ds \phi_i (x) := (\lambda I -
\Kcal_B^*)^{-1} (\frac{1}{i!} \nu_y \cdot \nabla y^i)(x), \quad x
\in \p B. $$


\section{Derivation of the Full Asymptotic Formula}


In this section we derive our asymptotic formula (\ref{final3}).
As stated in the introduction, we restrict our derivation to the
case of a single inhomogeneity ($m=1$).  We only give the details
when considering the difference between the fields  corresponding
to one and zero inhomogeneities. In order to further simplify
notation we assume that the single inhomogeneity $D$ has the form
 $D= \ep
B +z$, where  $z \in \Omega$ and $B$ is a bounded Lipschitz domain
in $\RR^d$ containing the origin. Suppose that the conductivity of
$D$ is $k$. Let $\lambda := \frac{k+1}{2(k-1)}$. Then by
\eqnref{KS} and \eqnref{KS-2}, the solution $u$ of (\ref{ud}) takes
the form
$$
u(x) = U(x) - N_D (\lambda I - \Kcal_D^*)^{-1}
(\pd{H}{\nu}|_{\p D})(x), \quad x \in \p\Om,
$$
where $U$ is the background potential given in \eqnref{back}.

  Define
$$ H_n(x) := \ds \sum_{|i| =0}^n \frac{1}{i!} (\p^i H)(z) (x-z)^i.
$$ Then we have from (\ref{KS-3}) that
\begin{align*}
\left \|\pd{H}{\nu} - \pd{H_n}{\nu} \right \|_{L^2 (\p D)} & \le
\sup_{x \in \p D} | \nabla H(x) - \nabla H_n(x) | |\p D|^{1/2} \\
& \le \| H \|_{\mathcal{C}^{n+1} (\overline D)} |x- z|^n |\p
D|^{1/2} \\
& \le C \| g \|_{L^2(\p \Om)} \ep^n |\p D|^{1/2} .
\end{align*}

 Note that
 \begin{equation}
 \label{ih2p}
 \mbox{if } \int_{\p D} h
d \sigma =0, \mbox{ then } \int_{\p D} (\lambda I -
\Kcal_D^*)^{-1} h d \sigma =0.
\end{equation}
If $\int_{\p D} h
d \sigma =0$, then we have for $x \in \partial \Omega$ that
\begin{align*}
\left | N_D (\lambda I - \Kcal_D^*)^{-1} h (x)
\right |
&= \left | \int_{\p D} [N(x-y)
- N (x-z)] (\lambda I - \Kcal_D^*)^{-1} h (y) d\sigma (y) \right | \\
& \le C \ep |\p D|^{1/2} \left \| h  \right \|_{L^2 (\p D)} .
\end{align*}
It then  follows that
\begin{align*}
\left | N_D (\lambda I - \Kcal_D^*)^{-1} (\pd{H}{\nu}|_{\p D}
- \pd{H_n}{\nu}|_{\p D})(x) \right | & \le C \ep |\p D|^{1/2}
\left \|\pd{H}{\nu} - \pd{H_n}{\nu} \right \|_{L^2 (\p D)} \\
& \le C \| g \|_{L^2(\p \Om)} \ep^{d+n}.
\end{align*}
Therefore,  we have
\begin{equation} \label{asym-10}
u(x) = U(x) - N_D
(\lambda I - \Kcal_D^*)^{-1} (\pd{H_n}{\nu}|_{\p D})(x) +
O(\ep^{d+n}), \quad x \in \p\Om
\end{equation}
where $O(\ep^{d+n})$ term is dominated by $ C \| g \|_{L^2(\p \Om)} \ep^{d+n}$
for some $C$ depending only on $c_0$.
Note that
$$
(\lambda I - \Kcal_D^*)^{-1}
(\pd{H_n}{\nu}|_{\p D})(x) =  \sum_{|i| =1}^n (\p^i H)(z)
(\lambda I - \Kcal_D^*)^{-1} (\frac{1}{i!} \nu_x \cdot \nabla
(x-z)^i) (x).
$$
Since $D= \ep B + z$, one can prove by using the
change of
 variables $y = \frac{x-z}{\epsilon}$
that
$$
 (\lambda I - \Kcal_D^*)^{-1} (\frac{1}{i!} \nu_x
\cdot \nabla (x-z)^i) (x)
 = \ep^{|i|-1} (\lambda I -
\Kcal_B^*)^{-1} (\frac{1}{i!} \nu_y \cdot \nabla y^i)
(\frac{1}{\ep} (x-z)).
$$
Put
\begin{equation}
\label{phi} \ds \phi_i (x) := (\lambda I - \Kcal_B^*)^{-1}
(\frac{1}{i!} \nu_y \cdot \nabla y^i)(x), \quad x \in \p B.
\end{equation}
Then we get
\begin{align}
& N_D (\lambda I - \Kcal_D^*)^{-1} (\pd{H_n}{\nu}|_{\p D})(x) \label{asym-20} \\
& = \sum_{|i| =1}^n (\p^i H)(z)  \ep^{|i|-1} \int_{\p D} N(x,y)
\phi_i (\ep^{-1} (y-z)) d \sigma (y) \nonumber \\
& = \sum_{|i| =1}^n (\p^i H)(z)  \ep^{|i|+d-2 } \int_{\p B} N(x, \ep y + z )
\phi_i (y) d \sigma (y) . \nonumber
\end{align}

We now expand $N(x, \ep y + z )$ asymptotically as $\ep \to 0$. By
\eqnref{Neumann-2} we have the following relation: $$ (-
\frac{1}{2} I + \Kcal_\Om ) [N( \cdot, \ep y + z)](x) = \Gamma
(x-z- \ep y) \quad \mbox{modulo constants}, \quad x \in \p \Om. $$
Since $$ \Gamma(x - \ep y) = \sum_{|j|=0}^{+\infty} \frac{1}{j!}
\epsilon^{|j|} \p^j (\Gamma(x)) y^j, $$ we obtain
\begin{align*}
(- \frac{1}{2} I + \Kcal_\Om ) [N( \cdot, \ep y + z)](x)
& = \sum_{|j|=0}^{+\infty} \frac{1}{j!} \epsilon^{|j|}
\p^j (\Gamma(x-z)) y^j \\
& = (- \frac{1}{2} I + \Kcal_\Om ) \left [
 \sum_{|j|=0}^{+\infty} \frac{1}{j!} \epsilon^{|j|}
\p^j_z N (\cdot, z) y^j \right ] (x).
\end{align*}
Since $\int_{\p\Om} N(x,w) d \sigma(x) =0$ for all $w \in \Om$,
we have the following asymptotic expansion of the Neumann function which is
of independent interest.

\begin{lem} \label{exp-n}
For $x \in \p\Om$, $z \in \Om$, and $y \in \p B$, and $\ep \to 0$,
\begin{equation} \label{exp-n1}
N(x, \ep y + z) =  \sum_{|j|=0}^{+\infty} \frac{1}{j!} \epsilon^{|j|}
\p^j_z N (x, z) y^j .
\end{equation}
\end{lem}

We now have from \eqnref{asym-20}
\begin{align*}
& N_D (\lambda I - \Kcal_D^*)^{-1} (\pd{H_n}{\nu}|_{\p D})(x)  \\
& = \sum_{|i| =1}^n (\p^i H)(z)  \ep^{|i|+d-2 }
 \sum_{|j|=0}^{+\infty} \frac{1}{j!} \epsilon^{|j|}
\p^j_z N (x, z) \int_{\p B}  y^j
\phi_i (y) d \sigma (y) .
\end{align*}
Observe that $$ \sum_{|i| = l} \frac{1}{i!} (\p^i H)(z)
\Delta(y^i) = \Delta_y \left ( \sum_{|i| = l} \frac{1}{i!} (\p^i
H)(z) y^i \right ) =0, $$ and therefore, by the Green's theorem,
it follows that $$\ds \int_{\partial B} \sum_{|i| = l}
\frac{1}{i!} (\p^i H)(z) \nabla (y^i)\cdot \nu(y) \; d\sigma(y) =
0.$$ Thus, in view of (\ref{phi}), the following identity holds by
using observation (\ref{ih2p})
\begin{equation} \label{ih}
\sum_{|i| = l} (\partial^i H)(z) \int_{\p B} \phi_i (y) d
\sigma(y)=0, \quad \forall \; l \geq 1.
\end{equation}
In fact, this follows immediately from (\ref{ih2p}).
Recall now that $\ds M_{ij} = \int_{\p B} y^j  \phi_i (y) d
\sigma(y)$ is the generalized polarization tensor associated with
the domain $B$ and the conductivity $k$  to obtain the following
pointwise asymptotic formula: for $x \in \p\Om$,
\begin{equation} \label{3600}
u(x) = U(x) -  \ep^{d-2} \sum_{|i| =1}^n \sum_{|j|=1}^{n-|i|
+1} \frac{1}{j!} \epsilon^{|i| + |j|}
 (\p^i H)(z) M_{ij} \p^j_z N(x,z)  +
O(\ep^{d+n}).
\end{equation}

Observing that the formula \eqnref{3600} still contains $\p^i H$ factors,
the remaining task is to convert \eqnref{3600} to a formula given solely by $U$ and its derivatives.

As a simpliest case, let us now take $n=1$ to find the leading order term in the
asymptotic expansion of $u|_{\partial \Omega}$ as
$\epsilon \rightarrow 0$. From \eqnref{KS} and \eqnref{rep-n-1},
we get
$$
\| H - U \|_{L^\infty (\p \Om)} \le C \ep^d \| \phi \|_{L^2(\p D)}
\le C \ep^d \| g \|_{L^2(\p\Om)}
$$
for some $C$ depending only on $\Om$ and $c_0$. It then follows
from the maximum principle that
$$
\| H - U \|_{L^\infty (\Om)} \le C \ep^d \| g \|_{L^2(\p\Om)}.
$$
Then, from the mean value property of harmonic functions, we
obtain
$$
|\nabla H(z) - \nabla U(z)| \le C \ep^d \| g \|_{L^2(\p\Om)}.
$$
It thus follows from \eqnref{3600} that
\begin{equation} \label{for2}
u(x) = U(x) - \ep^d \sum_{|i| =1, |j|=1} (\p^i U)(z) M_{ij} \p^j
N(x,z) + O(\ep^{d+1}), \quad x \in \p\Om,
 \end{equation}
which is in view of (\ref{polar}) exactly the formula derived in
\cite{FV89} and \cite{CMV98} when $D$ has $C^{1, \alpha}$
boundary.

\medskip

We now return to (\ref{3600}). Substitution of \eqnref{3600} into
\eqnref{KS-1} yields that,
for any $x \in \Om$,
\begin{equation}
\label{h1g}
 H(x) = U(x) -  \ep^{d-2} \sum_{|i| =1}^n
\sum_{|j|=1}^{n-|i| +1} \frac{1}{j!} \epsilon^{|i| + |j| } ( \p^i
H)(z) M_{ij} \Dcal_\Omega ( \p^j_z N(\cdot ,z))(x) + O(\ep^{d+n}).
\end{equation}
In (\ref{h1g}) the remainder $O(\ep^{d+n})$ is uniform in the
$\mathcal{C}^n$ norm on any compact subset of $\Om$ for any $n$
and therefore
\begin{equation}
 (\p^l H)(z) + \sum_{|i| =1}^n
\ep^{d-2} \sum_{|j|=1}^{n-|i| +1} \epsilon^{|i| + |j| } (\p^i
H)(z) P_{ijl} = (\p^l U)(z) + O(\ep^{d+n}),
\end{equation}
for all $l \in \NN^d$ with $|l| \leq n$ where
\begin{equation} \label{pijl}
P_{ijl} =
\frac{1}{j!} M_{ij} \p^l_x \Dcal_\Omega (\p^j_z N(\cdot ,
z))|_{x=z} .
\end{equation}
Define the operator
$$
\Pcal_\epsilon :
(v_l)_{l\in \NN^d, |l| \leq n} \mapsto \Bigr ( v_l + \ep^{d-2}
\sum_{|i| =1}^n \sum_{|j|=1}^{n-|i| +1} \epsilon^{|i| + |j|} v_i
P_{ijl} \Bigr )_{l\in \NN^d, |l| \leq n} .
$$
Observe that
$$
\Pcal_\epsilon = I + \epsilon^d \Pcal_1 + \ldots +
\epsilon^{n+d-1} \Pcal_{n-1} .
$$
Defining $ \Qcal_p, p=1, \ldots,
n-1,$ by
\begin{equation} \label{qcal}
(I + \epsilon^d \Pcal_1 +
\ldots + \epsilon^{n+d-1} \Pcal_{n-1})^{-1} = I + \epsilon^d
\Qcal_1 + \ldots + \epsilon^{n+d-1} \Qcal_{n-1} +
O(\epsilon^{n+d}),
\end{equation}
we finally obtain that
\begin{equation}
 ((\p^i H)(z))_{i\in \NN^d, |i| \leq n}  =
 (I + \sum_{p=1}^{n} \epsilon^{d + p-1}
  \Qcal_p) ((\p^i U)(z))_{i\in \NN^d, |i| \leq n}
   + O(\ep^{d+n})
\end{equation}
which yields the main result of this paper stated in Theorem
\ref{thm:main}.

\medskip

We also have a complete asymptotic expansion of the solutions of the Dirichlet problem:

\begin{thm} \label{thm:main2}
Suppose that the inhomogeneity consist of single component and let
$u$ be the solution of \eqnref{nepsg} with the Neumann condition
replaced by the Dirichlet condition $u|_{\p\Om} =f$. Let $V$ be
the solution of $\Delta V=0$ in $\Om$ with $V|_{\p\Om}=f$. The
following pointwise asymptotic expansion on $\partial \Omega$
holds for $d=2,3$:
\begin{equation} \label{final4}
\begin{array}{ll}
\ds \pd{u}{\nu} (x) &= \ds \pd{V}{\nu} (x) -  \ds  \ep^{d-2}
\sum_{|i| =1}^n \sum_{|j|=1}^{n-|i| +1} \frac{1}{j!} \epsilon^{|i|
+ |j|} \times \\ \nm & \quad \quad \ds \Bigr[ \ds \Bigr( (I +
\sum_{p=1}^{n +2 - |i| -|j| -d} \epsilon^{d + p-1}
  \Qcal_p) (\partial^l U(z)) \Bigr)_i  M_{ij} \p^j_z \pd{}{\nu_x}G(x,z) \Bigr]
\\
\nm &  \quad + \ds  O(\ep^{d+n}),
\end{array}
\end{equation}
where the remainders $ O(\ep^{d+n})$ are is dominated by $C
\ep^{d+n} \| g \|_{L^2(\p \Om)}$ for some $C$ independent of $x\in
\p\Om$. Here $G(x, z)$ is the Dirichlet Green function, $M_{ij}$,
$i,j \in \NN^d$,  are the generalized polarization tensors, and
$\Qcal_p$ is the operator defined in \eqnref{qcal} where
$\Pcal_{ijk}$ is defined, in this case, by
\begin{equation} \label{pijl-1}
P_{ijl} = \frac{1}{j!} M_{ij} \p^l_x \Scal_\Omega (\p^j_z
(\pd{}{\nu_x} G) (\cdot , z))|_{x=z} .
\end{equation}
\end{thm}

Theorem \ref{thm:main2} can be proved in the exactly same manner
as Theorem \ref{thm:main}. We begin with Lemma \ref{rep-d}. Then
the same arguments give us
\begin{equation*}
u(x) = U(x) -  \ep^{d-2} \sum_{|i| =1}^n \sum_{|j|=1}^{n-|i|
+1} \frac{1}{j!} \epsilon^{|i| + |j|}
 (\p^i H)(z) M_{ij} \p^j_z  G(x,z)  +
O(\ep^{d+n}).
\end{equation*}
From this we can get \eqnref{final4} as before.

\medskip
We conclude this paper by making a remark.  The
following formula is not exactly an asymptotic formula. However,
since the formula is simple and has some potential applicability
in solving the inverse conductivity problem, we make a record of
it as a theorem.

\begin{thm} \label{thm:H}
We have
\begin{equation} \label{for1}
u(x) = H(x) + \ep^{d-2} \sum_{|i| =1}^n \sum_{|j|=1}^{n-|i| +1}
\frac{1}{j!} \epsilon^{|i| + |j|}  \p^i H(z) M_{ij}
\p^j \Gamma (x-z) +  O(\ep^{d+n}),
\end{equation}
where $x \in \Om_0$ and $O(\ep^{d+n})$ term is dominated by $ C \| g \|_{L^2(\p \Om)} \ep^{d+n}$
for some $C$ depending only on $c_0$, and $H$ is given in (\ref{KS-1}).
\end{thm}


\end{document}